\documentclass[aps,pre,twocolumn, superscriptaddress,showpacs,longbibliography]{revtex4-2}
\usepackage{dcolumn,bm,graphics,epsfig,amsmath,amssymb,color,graphicx,amsthm,amssymb,verbatim,xcolor}
\usepackage{hyperref}
\definecolor{brickred}{rgb}{0.7, 0.25, 0.33}

\newcommand{\be}{\begin{equation}}
\newcommand{\ee}{\end{equation}}

\def\bc{\begin{center}}
\def\ec{\end{center}}
\newcommand{\avg}[1]{\langle{#1}\rangle}
\newcommand{\Avg}[1]{\left\langle{#1}\right\rangle}

\begin{document}

\title{$(k,n)$-core percolation on hypergraphs with anchor nodes}

\author{Hoseung Jang}
\affiliation{Advanced-Basic-Convergence Research Institute, Chungbuk National University, Cheongju, Chungbuk 28644, Korea}
\author{Byungjoon Min}
\email{bmin@cbnu.ac.kr}
\affiliation{Advanced-Basic-Convergence Research Institute, Chungbuk National University, Cheongju, Chungbuk 28644, Korea}
\affiliation{Department of Physics, Chungbuk National University, Cheongju, Chungbuk 28644, Korea}
\author{Ginestra Bianconi}
\email{ginestra.bianconi@gmail.com }
\affiliation{Centre for Complex Systems, School of Mathematical Sciences, Queen Mary University of London, Mile End Road, London, E1 4NS, UK}

\begin{abstract}
Hypergraphs describe higher-order interactions that involve more than a pair of nodes.
A characteristic feature of hypergraphs is that their robustness can be strongly affected by the different roles of the nodes. Indeed, some nodes might be essential for a hyperedge's function, while others might not be. The loss of a single essential node completely destroys the hyperedge it belongs to, while the  loss of a non-essential node  has a buffering effect,  inducing the hyperedge to simply reduce its size. In order to capture this phenomenology, we formulate a comprehensive theoretical framework for  $(k,n)$-core percolation models on hypergraphs, 
where each node of a hyperedge is an anchor with probability $\theta$, and 
a hyperedge fails if an anchor node fails. Hypergraph $(k,n)$-core percolation problems can be classified as first-neighbor and second-neighbor problems, indicating that in the pruning process the connectivity is ensured only by the state of the first neighbors or the second neighbors, respectively.  We derive self-consistency equations for first-neighbor and second-neighbor 
(node- and hyperedge-based) pruning processes, and obtain the size of the giant $(k,n)$-core.
We obtain the phase diagram, including continuous and discontinuous transitions, and confirm our theory on random hypergraphs using numerical simulations.
The results show how the heterogeneity of the nodes' functional roles and the extended range of the  interactions affect the robustness of higher-order networks. 
\end{abstract}

\maketitle

\section{Introduction}

Many complex systems include higher-order interactions that involve more than two nodes \cite{bianconi2021higher,battiston2020networks,battiston2021physics,perc2022dynamics}. 
Higher-order networks including simplicial complexes and hypergraphs naturally represent such interactions and appear as the ideal representation for a wide variety of systems, from biological to social and technological applications \cite{bianconi2021higher,boccaletti2023structure}.
This has motivated growing work on higher-order random graph models \cite{courtney2016generalized,barthelemy2022class},
contagion processes \cite{iacopini2019simplicial,de2020social,ferraz2024contagion,kim2024higher,burgio2024triadic},
spin models \cite{robiglio2025higher,son2026phase},
synchronization \cite{skardal2019abrupt,millan2020explosive}, percolation \cite{sun2023dynamic,bianconi2024theory},
and opinion formation \cite{neuhauser2020multibody,horstmeyer2020adaptive,kim2025competition}.
Consequently, characterizing the principles governing higher-order networks has emerged as 
a central theme in statistical physics and network science \cite{bianconi2021higher,battiston2021physics}.

An important aspect of this growing body of work is the realization that including  higher-order interactions leads to collective phenomena due to the interplay between structure and dynamics that have no equivalence in the context of pairwise networks \cite{millan2025topology,abiad2026hypergraphs}. Specifically, in percolation, the robustness of a hypergraph can be strongly affected by the different roles of the nodes \cite{bianconi2024theory,shang2025percolation,sun2026directionality}.   Indeed, nodes can be essential or non-essential for a hyperedge. The failure of an essential node completely dismantles the hyperedges to which it belongs, while the failure of a non-essential node induces only a reduction of the size of the hyperedges to which it belongs. For instance, in an online chat group, if a participant leaves the group, typically the chat  group simply decreases its size, however, it might occur that in a team an individual  plays  such a key role that their withdrawal might affect the entire performance of the team. Moreover, in biological networks, if a core catalytic subunit fails, the entire complex stops functioning.  
Similarly, in industrial supply chains, a manufacturing project depends on various suppliers. The loss of an ordinary component supplier can be tolerated as long as enough other suppliers remain active. 
In contrast, if essential suppliers fail, the entire production stops  immediately. From these examples, it is clear that the rich interplay between the higher-order structure and the functional role  of  the nodes can have  dramatic effects on the robustness of hypergraphs.

The $k$-core percolation \cite{dorogovtsev2006k,carmi2007model,cellai2011tricritical,baxter2011heterogeneous,hebert2013percolation,azimi2014k,baxter2015critical,cho2026recent}  results from a  pruning process in which nodes (or hyperedges) that fail to satisfy a minimum connectivity requirement are recursively removed. Recently, the extension of  this process from ordinary networks to hypergraphs  has raised significant interest  \cite{lee2023k,mancastroppa2023hyper,bianconi2024nature,zhao2025heterogeneous}.
Lee et al. \cite{lee2023k} treated all nodes as non-essential and introduced the 
$(k,n)$-core decomposition of hypergraphs, defined as the maximal subhypergraph 
in which every node has hypergraph degree at least $k$ and every hyperedge contains at least $n$ nodes. 
Bianconi and Dorogovtsev \cite{bianconi2024nature} treated all nodes as essential, and developed 
a message-passing theory of $k$-core percolation on hypergraphs, 
under the assumption that a hyperedge can be intact only if all of its incident nodes are intact. They thus revealed the difference in the robustness of hypergraphs when essential nodes are present.
Alongside these developments, hypergraph robustness and decomposition
have been explored through real hypergraph analysis and 
the formulation of hypergraph centrality measures \cite{mancastroppa2023hyper,mancastroppa2024structural,zhang2025accelerating} as well as other generalized $k$-core percolation problems \cite{zhao2025heterogeneous}.

In both of these formulations, every node in a hyperedge plays the same role in the hyperedge's function.
This is a reasonable approximation in some cases, but in some real higher-order networks 
the nodes participating in  higher-order interactions are essential with respect to a given hyperedge only with probability $\theta$ \cite{shang2025percolation,sun2026directionality}.  In this heterogeneous scenario, essential nodes are called anchor nodes~\cite{shang2025percolation,sun2026directionality}.
Incorporating such critical roles into higher-order network theory is important, 
since previous $k$-core percolation models on hypergraphs assume that all nodes within a hyperedge are functionally homogeneous. 

To capture this heterogeneity, in this work, we propose a comprehensive theoretical framework for  $(k,n)$-core percolation models on hypergraphs with anchor nodes. 
Specifically, for each hyperedge, we draw independently at random its anchor nodes: each node of the hyperedge becomes one of its anchor nodes with  probability $\theta$. 
If an anchor node fails, the hyperedge fails regardless of the state of its remaining nodes.
In this setting, we do not only explore the robustness of hypergraphs when the connectivity is ensured  by the first neighbors in the factor graph representing the hypergraph, but  we also provide a comprehensive theoretical framework for second-neighbor $(k,n)$-core percolation problems.
The second-neighbor problems  allow us to explore the role of long-range connectivity, which is attracting considerable attention for network percolation problems \cite{cirigliano2023extended,cirigliano2024general,kim2024shortest,meng2025path,hu2025unveiling},  in the context of hypergraphs.

In the first-neighbor problems, a hyperedge with functioning anchor nodes remains active if a sufficient number $n$ of its nodes are connected to the giant component. 
While $n$ sets the minimum number of nodes that a hyperedge must retain in the giant component to remain active, the parameter $k$ sets the minimum number of active hyperedges a node must belong to, in order to remain in the giant component itself. 
Thus, in the first-neighbor problems, the $(k,n)$-core is obtained by the joint recursive pruning of nodes with fewer than $k$ active incident hyperedges and hyperedges with fewer than $n$ active nodes until no further removal is possible. 
The second-neighbor problems can be node- or hyperedge-based. In the node-based second-neighbor problems, the connectivity of a given node  depends only on the state of the nodes that share a hyperedge  with it. In the hyperedge-based second-neighbor problems, the connectivity of a given hyperedge  depends only on the state of the hyperedges that share a node  with it. 
By providing a comprehensive theoretical framework that encompasses both first-neighbor and second-neighbor $(k,n)$-core percolation problems, we are able to assess the role of extended-range connectivity in the context of hypergraphs with anchor nodes in determining the robustness of hypergraphs and their phase diagram.

This paper is organized as follows. 
In Sec. II, we introduce the $(k,n)$-core percolation models on hypergraphs with anchor nodes. In Sec. III, we derive the self-consistency equations for the first-neighbor case and we study the behavior of the model. 
In Sec. IV, we develop the theory for the second-neighbor pruning process acting on nodes and hyperedges.
In both cases, we validate our analytical predictions against Monte-Carlo simulations on random hypergraphs.
Finally, in Sec. V, we summarize our results and discuss their implications for real-world higher-order 
systems with heterogeneous node dependencies.

\section{Hypergraphs $(k,n)$-core percolation with anchor nodes}

We consider a hypergraph $\mathcal{H} = (\mathcal{V}, \mathcal{E})$, where $\mathcal{V}$ is the set of $N$ nodes and $\mathcal{E}$ is the set of $M$ hyperedges. 
A hyperedge $e \in \mathcal{E}$ of size $m$ represents a higher-order interaction involving $m$ nodes. 
Nodes in a hyperedge can independently act as anchor nodes with probability $\theta$, 
and non-anchor nodes with probability $1-\theta$. 
Anchor nodes are essential to the functionality of the hyperedge: if an anchor node fails (removed or damaged), the hyperedge to which the anchor node belongs fails automatically, regardless of the states of the other nodes. 
Conversely, non-anchor nodes contribute to maintaining the hyperedge through a minimum connectivity requirement $n$.

The $(k,n)$-core of a hypergraph with anchor nodes is defined as follows. 
Each node in a hyperedge is independently designated as an anchor node of that hyperedge with probability $\theta$. 
A hyperedge is said to be active if it is undamaged, all of its anchor nodes are active, 
and at least $n$ of its nodes overall are active. 
A node is said to be active if it is undamaged and belongs to at least $k$ active hyperedges. 
The $(k,n)$-core is the maximal subhypergraph consisting of all active nodes and active hyperedges 
satisfying these conditions simultaneously.
By introducing the anchor nodes with probability $\theta$, our model captures this functional heterogeneity. 
If an anchor node is damaged (or removed), all hyperedges containing it are also removed.
Tuning $\theta$ allows us to study a wide range of $k$-core problems on hypergraphs. When $\theta = 0$, 
the model recovers the $(k,n)$-hypergraph percolation where all nodes are equal. 
As $\theta$ increases, the system becomes more dependent on essential components. 
This simple mechanism gives us a general framework to analyze robustness in higher-order networks.

It is useful to formulate the model on the factor graph of the hypergraph, 
i.e., the bipartite network whose two sets of nodes are the $N$ nodes and the $M$ hyperedges, 
a node $i$ and a factor node $\alpha$ being linked when $i\in\alpha$ [see Fig.~1(a)]. 
In the factor graph, the anchor property belongs to the links: 
each link is an anchor link with probability $\theta$. 
A node can therefore be an anchor of one group and an ordinary member of another.
Then, a factor node is removed if any of its anchor links points to a removed node, or if fewer than $n$ of its links point to surviving nodes, while a node is removed if fewer than $k$ of its links point to surviving factor nodes.

\begin{figure}
\includegraphics[width=\linewidth]{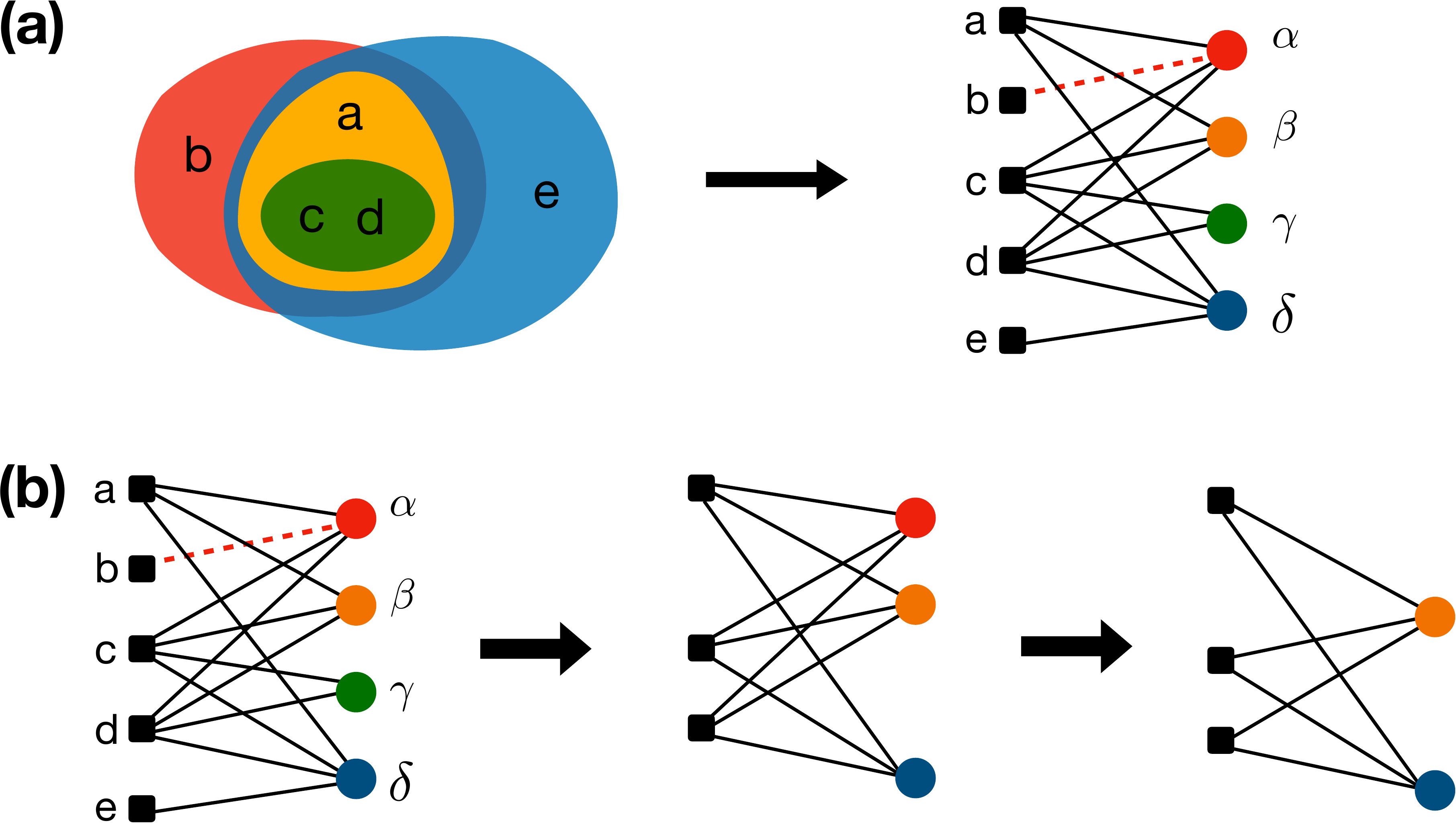}
\caption{
(a) Schematic illustration of the mapping between a hypergraph and its corresponding factor network. 
(b) An example of $(k,n)=(2,3)$-core first-neighbor pruning.
The nodes $b$ and $e$ are removed because they have fewer than $2$ hyperedges and the hyperedge $\gamma$ is removed because it has fewer than $3$ nodes. 
In addition, the hyperedge $\alpha$ is further removed due to the removed anchor node $b$, even though it has three active nodes.
}
\label{fig:fig1}
\end{figure}

\section{First-neighbor pruning}

The $(k,n)$-core can be identified by the following recursive pruning process. 
Starting from the full hypergraph, each hyperedge is independently damaged (removed) with probability $1-p_H$, 
and each node is independently damaged with probability $1-p_N$. 
All undamaged nodes and hyperedges are initially marked as active.
The following rules are then applied repeatedly until no further node or hyperedge changes state:
\begin{enumerate}
\item[(i)] A hyperedge becomes inactive if it is damaged, if any of its anchor nodes is inactive, or if it has fewer than $n$ active nodes.
\item[(ii)] A node becomes inactive if it is damaged, or if it has fewer than $k$ active hyperedges.
\item[(iii)] These two processes proceed, in any order, until there are no further newly deactivated entities.
\end{enumerate}
The nodes and hyperedges that remain active at the end of this iterative procedure 
constitute the $(k,n)$-core of the hypergraph with anchor nodes.
We refer to this procedure as first-neighbor pruning 
because the state of each node or hyperedge is evaluated using the current state of its first neighbors on the factor graph. 
An example of the pruning process is depicted in Fig.~1(b).

\subsection{Analytical approaches}

We develop the analytical theory to determine the size of the $(k,n)$-core 
with anchor nodes under first-neighbor pruning, based on the cavity approach.  
Let us indicate by $W$ the probability that by following a random link of the factor graph departing from a hyperedge we reach a node that is connected to the giant component, 
and let us indicate by $V$ the probability that by following a random link of the factor graph departing from a node we reach a hyperedge in the giant component.
The equations for $W$ and $V$ on a locally tree-like structure are given by 
\begin{align}
\label{mes1}
W&=p_N\sum_{q\geq k}\frac{qP(q)}{\avg{q}}\sum_{s=k-1}^{q-1}\binom{q-1}{s} V^s(1-V)^{q-1-s},\\
\label{mes2}
V&=p_H\sum_{m\geq n}\frac{mQ(m)}{\Avg{m}}\sum_{p=0}^{m-1}\binom{m-1}{p}\theta^{p}(1-\theta)^{m-1-p} \\
&\times \sum_{s=\text{max}(0,n-1-p)}^{m-1-p}\binom{m-1-p}{s}W^{s+p}(1- W)^{m-1-s-p}, \nonumber
%&\times \sum_{s=\max(0,n-1-p)}^{m-1-p}
\end{align}
where $P(q)$ and $Q(m)$ are the degree and cardinality distributions, respectively.
The first equation represents the fact that the node reached by following a link of the hypergraph departing from a hyperedge belongs to the giant component if and only if: 
(i) the node is not damaged and (ii) the node is connected by at least  $k-1$ hyperedges connected to the giant component.
The second equation expresses the fact that the hyperedge reached by following a link of the hypergraph departing from a node belongs to the giant component if and only if: 
(i) the hyperedge  is not damaged, (ii) all its anchor nodes are connected to the giant component, and (iii) a total of at least $n-1$ nodes is connected to the giant component.
The equations (\ref{mes1}) and (\ref{mes2}) self-consistently determine the values of $W$ and $V$,
as each of the two variables is expressed purely in terms of the other. 
One can obtain the fixed point of $W$ and $V$ by iteratively solving Eqs.~(\ref{mes1}) and (\ref{mes2}) to converge to the solution corresponding to the giant 
$(k,n)$-core.

The marginal probabilities $S$ and $R$ that a node and a hyperedge, respectively, belong to the $(k,n)$-core are given by
\begin{align}
\label{marginal1}
S&=p_N\sum_{q\geq k}{P(q)}\sum_{s=k}^{q}\binom{q}{s} V^s(1-V)^{q-s}, \\
R&=p_H\sum_{m\geq n}{Q(m)}\sum_{p=0}^{m}\binom{m}{p}\theta^{p}(1-\theta)^{m-p} \\
&\times \sum_{s=\text{max}(0,n-p)}^{m-p}\binom{m-p}{s}W^{s+p}(1-W)^{m-s-p}. \nonumber 
%&\sum_{s=\max(0,n-p)}^{m-p}\binom{m-p}{s}W^{s+p}(1-W)^{m-s-p} \nonumber 
\end{align}
Here, $S$ is the probability that a node with the degree distribution $P(q)$ is undamaged and connected to at least $k$ active hyperedges, 
and $R$ is the probability that a hyperedge with cardinality distribution $Q(m)$ is undamaged, 
has all of its anchor nodes active, and has at least $n$ out of its $m$ nodes active. 
The size of the $(k,n)$-core is given by $S$ for the fraction of active nodes and by $R$ for the fraction of active hyperedges.

We consider two limiting cases: the case in which there are no anchor nodes, $\theta=0$, and the case in which all nodes are anchor nodes, $\theta=1$.
In the first case ($\theta=0$), the equations for $V$ and for $R$ become
\begin{align}
V&=p_H\sum_{m\geq n}\frac{mQ(m)}{\Avg{m}}\sum_{s=n-1}^{m-1}\binom{m-1}{s}W^{s}(1- W)^{m-1-s}, \label{eq:Ref9_V}\\
R&=p_H\sum_{m\geq n}{Q(m)}\sum_{s=n}^{m}\binom{m}{s}W^{s}(1-W)^{m-s}, \label{eq:Ref9_R}
\end{align}
for which we recover the $(k,n)$-core model of Ref. \cite{lee2023k}.
In the second case where $\theta=1$, the equations for $V$ and for $R$ become
\begin{align}
V&=p_H\sum_{m\geq n}\frac{mQ(m)}{\Avg{m}}W^{m-1},\\
R&=p_H\sum_{m\geq n}{Q(m)}W^{m}.
\end{align}
Therefore, we recover the $(k,n)$-core model of Ref. \cite{bianconi2024nature}.

\begin{figure*}
\includegraphics[width=\linewidth]{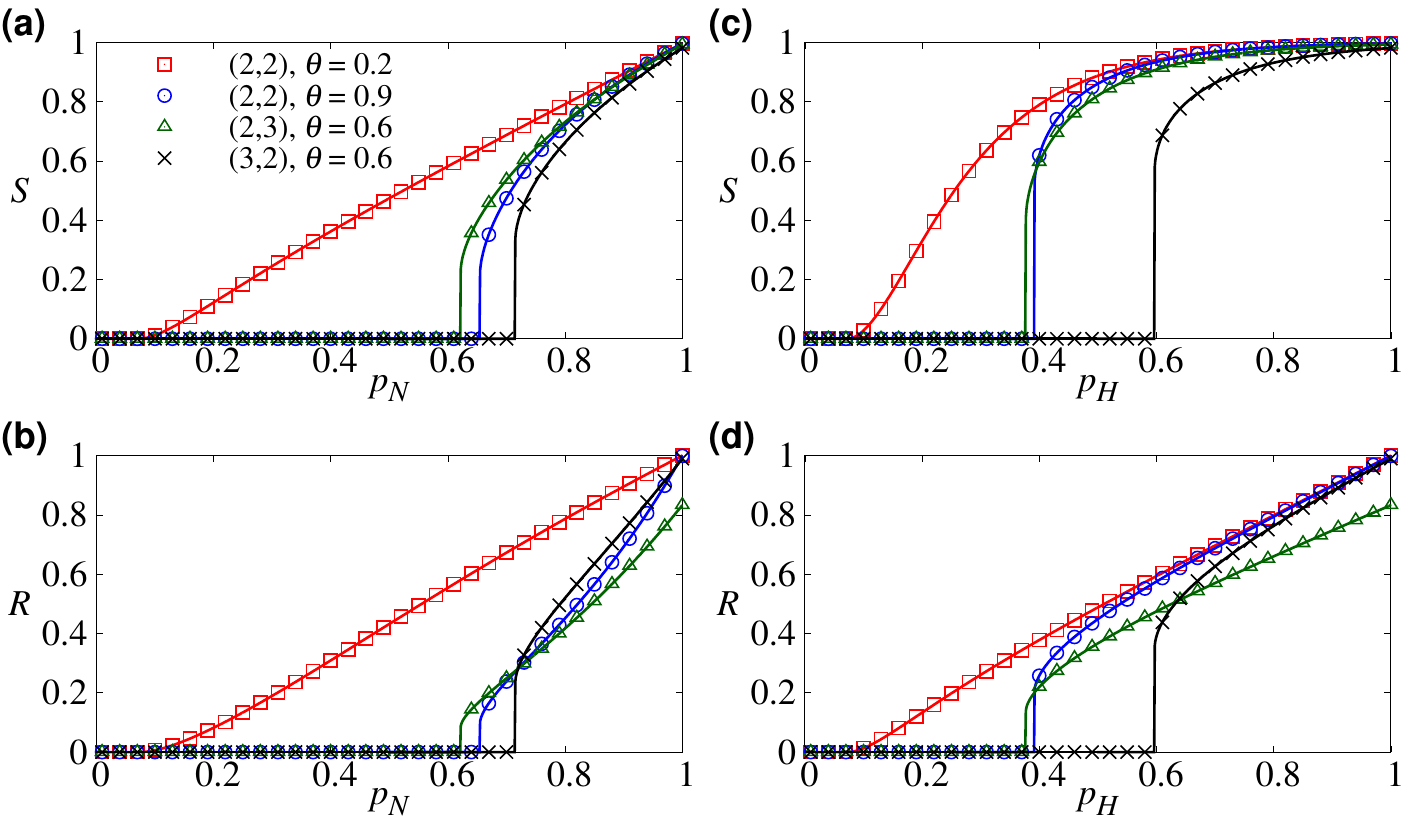}
\caption{
The size $S$ and $R$ as a function of (a,b) $p_N$ with $p_H=1$ and (c,d) $p_H$ with $p_N=1$ for various core thresholds $(k,n)$ and the anchor density $\theta$.
The numerical results (symbols) for $N=10^5$ and the analytical predictions (lines) are shown together.
All hypergraphs have Poisson degree and cardinality distributions given by Eqs.~(\ref{eq:Poisson}) and (\ref{eq:Shifted Poisson}) with $\langle q \rangle=7.6$ and $\langle m \rangle=3.8$.
}
\label{fig:fig2}
\end{figure*}

\subsection{Phase diagram}

The nature of the percolation transition is determined by the fixed points of Eqs.~(\ref{mes1}) and (\ref{mes2}). 
Since the self-consistency equations of $V$ and $W$ depend only on $W$ and $V$, respectively,  
we define them as $W=f(V), V=g(W)$.
The two equations can be decoupled by substitution, and the coupled equations reduce to the single equation $V=g(f(V))$.
Then, we define 
\begin{equation}
h(V)= V-g(f(V)),
\label{eq:h}
\end{equation}
whose roots $V^\star$ correspond to the fixed point.
Following the standard theory of critical phenomena, the transitions
are identified from the tangency conditions
\begin{equation}
 h(V^\star)=0,\quad h'(V^\star)=0, 
\label{eq:crit}
\end{equation}
where $h'(x)=dh(x)/dx$.
A continuous transition corresponds to $h'(0)=0$
and a discontinuous (hybrid) transition occurs at $V^{\star}>0$.
At the tricritical point, the two lines meet with the condition 
$h''(0)=0$, where $h''(x)=d^2 h(x) /dx^2$.

To see these transition criteria in detail, let us consider phase diagrams with specific parameters of $(k,n)$.
For the case $(k,n)=(2,2)$, the continuous transition occurs at
\begin{align}
p_N p_H\frac{\Avg{q(q-1)}_k}{\avg{q}}\frac{\Avg{m(m-1)(1-\theta)^{m-2}}_n}{\avg{m}}=1,
\end{align}
where $\Avg{h(q)}_k=\sum_{q\geq k}h(q)P(q)$ and $\Avg{h(m)}_n=\sum_{m\geq n}h(m)Q(m)$.
In addition, the tricritical point can be obtained for 
%\byu{[check]} \textcolor{red}{Checked: Solving $h'(0)=0 , h''(0)=0$ simultaneously}
\begin{align}
    &\frac{\avg{q}\avg{q(q-1)(q-2)}_k}{\avg{q(q-1)}_k^{2}}  \nonumber\\
&=p_N  (2\theta-1)  \frac{\avg{m(m-1)(m-2)(1-\theta)^{m-3}}_n}{\avg{m(m-1)(1-\theta)^{m-2}}_n}.
\label{eq:tri22}
\end{align}
When $k>2$, on the other hand, all transitions are discontinuous regardless of $\theta$.

\subsection{Results}

\begin{figure}
\includegraphics[width=\linewidth]{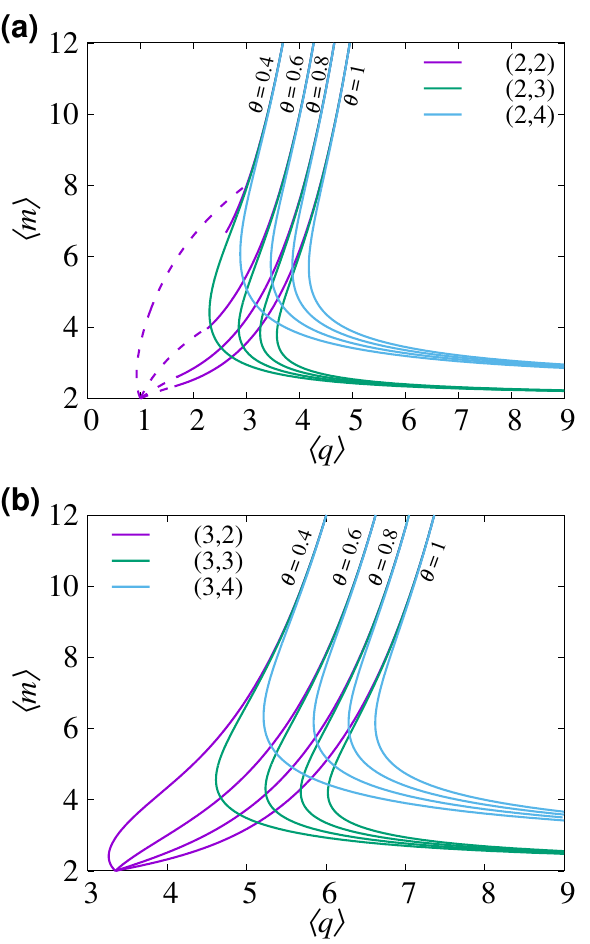}
\caption{
Phase diagram of $(k,n)$-core percolation with anchor nodes on random Poisson hypergraphs in the $\langle q \rangle$-$\langle m \rangle$ plane. Here, $p_N=p_H=1$.
The dashed lines and the solid lines denote continuous transitions and discontinuous transitions, respectively.
}
\label{fig:fig3}
\end{figure}

To validate the theory, we performed numerical simulations of the first-neighbor pruning process 
on random hypergraphs constructed by the configuration model with Poisson
degree distribution and shifted Poisson cardinality distribution:
\begin{align}
 P(q)&=e^{-\avg{q}}\frac{\avg{q}^{\,q}}{q!}, \label{eq:Poisson}\\
 Q(m\ge2)&=e^{-(\avg{m}-2)}\frac{(\avg{m}-2)^{m-2}}{(m-2)!}, \label{eq:Shifted Poisson}
 %\label{eq:Poisson}
\end{align}
with the shift being necessary because cardinalities $m=0,1$ are not allowed. 
Each node of each hyperedge is chosen to be an
anchor with probability $\theta$, nodes are damaged with
probability $1-p_N$ and hyperedges with probability $1-p_H$. 
In Fig.~\ref{fig:fig2}, we show the fraction of nodes $S$ and hyperedges $R$ belonging to the giant $(k, n)$-core as a function of the node survival probability $p_N$ [Figs.~\ref{fig:fig2}(a) and \ref{fig:fig2}(b)] and the hyperedge survival probability $p_H$ [Figs.~\ref{fig:fig2}(c) and \ref{fig:fig2}(d)] for various combinations of $(k, n)$ and anchor probability $\theta$ on random Poisson hypergraphs with $\langle q \rangle = 7.6$ and $\langle m \rangle = 3.8$. The analytical predictions (lines) obtained from the self-consistency equations are in excellent agreement with the Monte-Carlo simulations (symbols) across all parameter regimes.

Figure~\ref{fig:fig2}  shows that the nature of the percolation transition strongly depends on both the core thresholds $(k, n)$ and the anchor density $\theta$. For $(k, n) = (2, 2)$ with a low anchor density, e.g., $\theta = 0.2$, the giant core emerges continuously as $p_N$ or $p_H$ increases. 
In contrast, as the anchor density increases, i.e., $\theta = 0.9$ at the same $(k, n) = (2, 2)$, the transition turns to a discontinuous jump at a larger threshold $p_c$. This behavior indicates that an abundance of anchor nodes makes the hypergraph more fragile and the transition  changes from continuous to discontinuous. Furthermore, for higher core indices such as $(k, n) = (2, 3)$ and $(3, 2)$, the transitions are discontinuous regardless of $\theta$. 

We also provide the phase diagrams in the $\langle q \rangle$-$\langle m \rangle$ plane at $p_N = p_H = 1$ in Fig.~\ref{fig:fig3}. The phase boundaries separate the non-percolating phase (absence of the $(k, n)$-core) from the percolating phase. Figure~\ref{fig:fig3}(a) shows the phase boundaries for $k = 2$ with various $n =2,3,4$ for different values of $\theta$. For $(k, n) = (2, 2)$, the phase boundary is divided into a continuous transition line (dashed) and a discontinuous transition line (solid), meeting at the tricritical point. As $\theta$ increases, the continuous regime shrinks. In addition, for $n \ge 3$ [e.g., $(2,3)$ and $(2,4)$], the transition is purely discontinuous (solid curves) for all $\theta$.
Figure~\ref{fig:fig3}(b) shows the phase diagram for $k = 3$ with $n =2, 3, 4$. In these parameter values, all transitions are discontinuous. As either the hyperedge minimum requirement $n$ or the anchor probability $\theta$ increases, the hypergraph requires a denser structure (higher $\langle q \rangle$ and $\langle m \rangle$) to sustain a giant core. Consequently, the non-percolating region in the phase diagram expands with increasing $n$ and $\theta$. Therefore, higher anchor densities raise the connectivity threshold required for the emergence of the $(k, n)$-core.

\section{Second-neighbor pruning}

\subsection{Second-neighbor pruning of the nodes}

Under first-neighbor pruning, both nodes and hyperedges can be iteratively removed. 
An alternative scenario is one in which a hyperedge is not itself an entity that gets removed, 
but instead remains as long as its nodes remain connected. 
Only the nodes are recursively pruned, and a hyperedge can still belong to the giant component 
even if it contains fewer than $n$ active nodes, provided that 
its constituent nodes satisfy their $k$-core condition through other groups.
We call this the second-neighbor pruning of the nodes. 
The second-neighbor pruning of the nodes is obtained by the following procedure. 
Initially, nodes are damaged with probability $1-p_N$ and hyperedges with probability $1-p_H$, and then:
\begin{enumerate}
\item[(i)] A hyperedge becomes inactive if it is damaged or any of its anchor nodes is inactive.
\item[(ii)] A node becomes inactive if it is damaged, or if it belongs to
fewer than $k$ active hyperedges each connected to at least $n$ active nodes.
\item[(iii)] Rules (i) and (ii) are applied repeatedly until no further node changes state.
\end{enumerate}
The $(k,n)$-core is then the giant component induced by the active nodes and by the hyperedges attached to them, 
where a hyperedge belongs to the giant component if all of its anchor nodes are active and at least one of its nodes is active.

The coupled self-consistency equations, Eqs.~(\ref{mes1}) and (\ref{mes2}) governing the local probabilities $W$ and $V$ remain unchanged 
because the pruning of the nodes is identical to the first-neighbor case.
However, the expression for the order parameter, specifically the marginal fraction of hyperedges belonging to the giant component $R$, 
is modified to account for the presence of inactive hyperedges that still contribute to the core. 
Note also that the marginal size of the nodes $S$ remains unchanged.

To compute $R$, we classify the nodes of a hyperedge by how many other active hyperedges they belong to.
To this end we introduce two probabilities, $U$ and $\tilde{U}$, in which 
$U$ is the probability that a node reached by following a random hyperedge is connected to at least $k$ other active hyperedges, 
and $\tilde{U}$ is the probability that it is connected to exactly $k-1$ other active hyperedges. 
The probabilities are given by
\begin{align}
 U&=p_N\sum_{q \ge k+1}\frac{qP(q)}{\langle q \rangle}\sum_{s=k}^{q-1}\binom{q-1}{s} V^s(1-V)^{q-1-s}, \\
\tilde{U}&=p_N\sum_{q\geq k}\frac{qP(q)}{\Avg{q}}\binom{q-1}{k-1}V^{k-1}(1-V)^{q-k}.
\end{align}
The distinction between the two matters here: 
a node with the probability $U$ is active irrespective of the state of the hyperedge under consideration, 
whereas a node counted by $\tilde{U}$ is active if and only if that hyperedge is itself active.
Note that, by definition, $U+\tilde U=W$.

A hyperedge belongs to the giant component if all of its anchor nodes are
active and at least one of its nodes is active. 
Depending on the number of anchor nodes $p$ in a hyperedge, we classify three cases:
(i) If the hyperedge has at least one anchor node and is active, then all of its $p$ anchor nodes and $s$ of its non-anchor nodes send a positive message, with $s+p\ge n$, each contributing a factor $W$.  
(ii) If it has at least one anchor node but is inactive, so that $s+p<n$, its nodes counted by $\tilde U$, 
which are connected to exactly $k-1$ other active hyperedges, cannot satisfy the $k$-core condition. Thus, all $p$ anchor nodes must be connected to at least $k$ active hyperedges, contributing $U^p$.
%are no longer active, and its anchor nodes are required to be connected to at least $k$ other active hyperedges, contributing $U^p$. 
In both cases, the surviving anchor nodes guarantee that at least one node of the hyperedge is active.  
(iii) If the hyperedge has no anchor node at all, it fails to belong to the giant component only when none of its nodes is connected to at least $k$ other active hyperedges and fewer than $n$ of them are connected to exactly $k-1$. 
Taking these rules into account, the marginal probability $R$ that a randomly chosen hyperedge belongs to the giant component is given by
\begin{widetext}
\begin{align}
\label{eq:2ndr}
R&=p_H\left[\sum_{m}{Q(m)}\sum_{p=1}^{m}\binom{m}{p}\theta^{p}(1-\theta)^{m-p}\sum_{s=0}^{m-p}\binom{m-p}{s}W^{s+p}(1-W)^{m-s-p}H(s+p-n)\right]\\
&+p_H\left[\sum_{m}{Q(m)}\sum_{p=1}^{m}\binom{m}{p}\theta^{p}(1-\theta)^{m-p}U^p\sum_{s=0}^{m-p}\binom{m-p}{s}W^{s}(1-W)^{m-s-p}(1-H(s+p-n))\right]\nonumber\\
&+p_H \left[\sum_{m}{Q(m)}(1-\theta)^{m} \left(1-\sum_{s=0}^{\min(n-1,m)}\binom{m}{s}\tilde{U}^s(1-\tilde{U}-U)^{m-s}\right)\right], \nonumber
\end{align}
\end{widetext}
where $H(x)$ indicates the Heaviside function $H(x)=1$ if $x\geq 0$ and $H(x)=0$ if $x<0$.

We consider two special cases where $\theta = 1$ and $\theta =0$.
For $\theta=1$, every node is an anchor node and only the term $p=m$ of
Eq.~(\ref{eq:2ndr}) survives. 
A hyperedge of cardinality $m \ge n$ belongs to the giant component if all of its nodes belong to the core, 
while a hyperedge with $m<n$ can never be active, so that each of its nodes has to satisfy the $k$-core condition through other hyperedges alone.  
Hence,
\begin{equation}
R=p_H \left[\sum_{m<n}Q(m) U^m+\sum_{m\ge n} Q(m)W^m \right].
\label{eq:Rtheta1}
\end{equation}
For $\theta=0$, no node is an anchor node, and then 
\begin{equation}
%R=p_H\sum_m Q(m)\left[1-\sum_{s=0}^{\min(n-1,m)}\binom{m}{s}
R=p_H\sum_m Q(m)\left[1-\sum_{s=0}^{n-1}\binom{m}{s}
\tilde U^{s}(1-U-\tilde U)^{m-s} \right].
\label{eq:Rtheta0}
\end{equation}
In this case, a hyperedge belongs to the core if one of its nodes is active.

\subsection{Second-neighbor pruning of the hyperedges}

In the second-neighbor node-pruning scheme, only the nodes were pruned, whereas in the second-neighbor hyperedge-pruning process we consider instead that only the hyperedges are removed.
Interestingly, due to the presence of anchor nodes, there is no symmetry between these two second-neighbor pruning processes.
The second-neighbor pruning of the hyperedges consists of the  recursive removal of hyperedges,
while the nodes are not pruned and a node remains part of the core as long as it is attached to the surviving structure. 
Specifically, we repeatedly apply the following rules until no hyperedge changes state:
\begin{enumerate}
\item[(i)] The activity of the nodes is evaluated, a node becoming inactive if it belongs to fewer than $k$ surviving hyperedges.
\item[(ii)] A hyperedge is removed if it is damaged, if any of its anchor nodes is damaged, or if fewer than $n$ of its nodes are active.
\item[(iii)] Rules (i) and (ii) are repeated until no further hyperedges are removed.
\end{enumerate}
The $(k,n)$-core is the giant component induced by the surviving hyperedges,
together with the undamaged nodes attached to them.  
Note that the anchor condition now involves the initial damage of the anchor nodes rather than their activity.

Under this pruning process, the cavity equation for $W$ is unchanged from the first-neighbor case, since a node's $k$-core condition still depends on the same activity of its connected hyperedges, with the condition $p_N=1$. 
However, the equation for $V$ is modified because a hyperedge's activity depends on the hyperedges to which its nodes belong. 
The self-consistency equations are then given by
\begin{align}
W&=\sum_{q\geq k}\frac{qP(q)}{\avg{q}}\sum_{s=k-1}^{q-1}\binom{q-1}{s} V^s(1-V)^{q-1-s}, \\
V&=p_H\sum_{m\geq n}\frac{mQ(m)}{\langle m \rangle}\sum_{s=n-1}^{m-1}\binom{m-1}{s}A^{s}B^{m-1-s}, \label{eq:2nd_edge_V}
\end{align}
where
\begin{align}
A&=p_NW, \\
B&=\theta p_N(1-W)+(1-\theta)(1-p_NW) =\pi_N-A.
\end{align}
Here, $\pi_N=1-\theta(1-p_N)$.
In the above expressions, $A$ is the probability that a node is intact and connected to the giant component, and the probability $B$ instead groups together the nodes that are not connected to the giant component but do not destroy the hyperedge. 
The remaining term $1-\pi_N$ is the probability that a
node is a damaged anchor and therefore removes the hyperedge.  
Among the $m-1$ other nodes of a hyperedge, the hyperedge is active if at least $n-1$ are of type $A$, all the others being of type $B$.

The fraction of nodes $S$ is modified because the core contains not only the active nodes but also the undamaged nodes that are attached to at least one
surviving hyperedge.  
An undamaged node belongs to the core unless none of its hyperedges is connected to at least $n$ other active nodes 
and it belongs to fewer than $k$ hyperedges connected to exactly $n-1$ other active nodes. 
The auxiliary variables $\hat V$ and $\tilde V$ represent the probabilities of the two cases:
\begin{align}
\hat{V}&=p_H\sum_{m\geq n+1}\frac{mQ(m)}{\langle m \rangle}\sum_{s=n}^{m-1}\binom{m-1}{s}A^{s}B^{m-1-s}, \\
\tilde{V}&=p_H\sum_{m\geq n}\frac{mQ(m)}{\langle m \rangle}\binom{m-1}{n-1}A^{n-1}B^{m-n}.
\end{align}
Therefore, the marginal probabilities are given by
\begin{align}
\label{eq:2nd_edge}
R&=p_H\sum_{m\geq n}{Q(m)}\sum_{s=n}^{m}\binom{m}{s}A^{s}B^{m-s}, \\
S&=p_N\sum_{q}{P(q)}\left[1-\sum_{s\leq \min(k-1,q)}\binom{q}{s} \tilde{V}^s(1-\tilde{V}-\hat{V})^{q-s}\right].
\end{align}

For $\theta=0$, no node is an anchor node, $\pi_N=1$, and $B=1-p_N W$.
Then, Eqs.~(\ref{eq:2nd_edge_V}) and~(\ref{eq:2nd_edge}) reduce to
\begin{align}
    V&=p_H\sum_{m\ge n}\frac{mQ(m)}{\avg{m}}\sum_{s=n-1}^{m-1}\binom{m-1}{s}(p_NW)^s(1-p_NW)^{m-1-s}, \\
    R&=p_H\sum_{m\geq n}{Q(m)}\sum_{s=n}^{m}\binom{m}{s}(p_NW)^s(1-p_NW)^{m-1-s}.
\end{align}
Since $p_NW$ is identical to Eq.~(\ref{mes1}), these are equal to Eqs.~(\ref{eq:Ref9_V}) and~(\ref{eq:Ref9_R}), for which we recover the $(k,n)$-core model of Ref.~\cite{lee2023k}.
For $\theta=1$, every node is an anchor node, $\pi_N=p_N$, and
\begin{equation}
A^sB^{m-1-s}=p_N^{m-1}W^s(1-W)^{m-1-s} ,
\end{equation}
so that Eqs.~(\ref{eq:2nd_edge_V}) and~(\ref{eq:2nd_edge}) reduce to
\begin{align}
V&=p_H\sum_{m\ge n}p_N^{m-1}\frac{mQ(m)}{\avg{m}}\sum_{s=n-1}^{m-1}
   \binom{m-1}{s}W^s(1-W)^{m-1-s}, \\
R&=p_H\sum_{m\geq n}p_N^{m}Q(m)\sum_{s=n}^{m}\binom{m}{s}W^{s}(1-W)^{m-s},
\end{align}
which coincide with Ref.~\cite{bianconi2024nature}.

\subsection{Results}

\begin{figure}
\includegraphics[width=\linewidth]{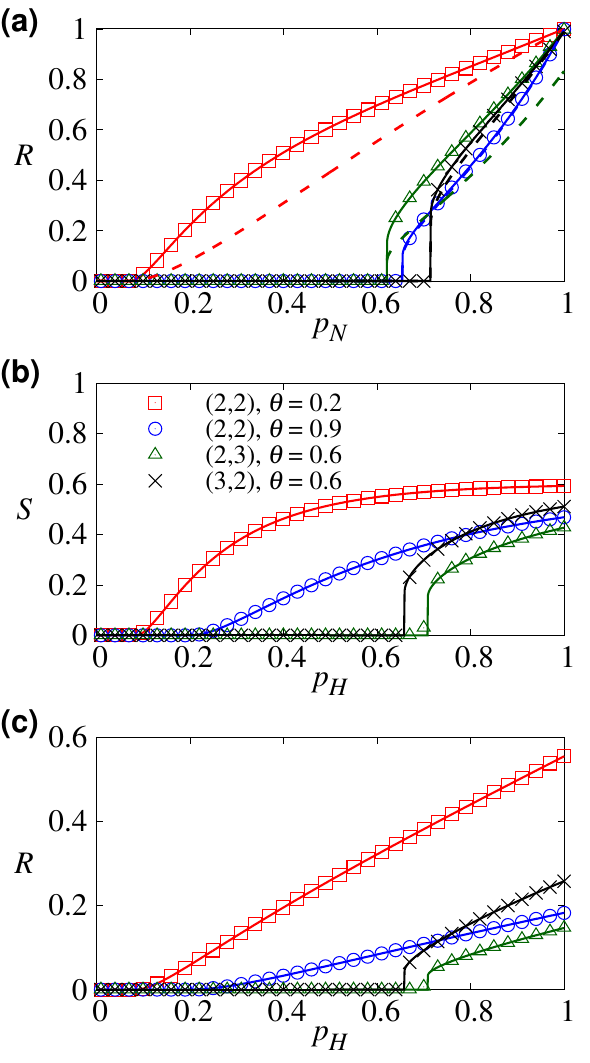}
\caption{
(a) The size $R$ as a function of $p_N$ for the second-neighbor node pruning with $p_H=1$.
The dashed lines correspond to $R$ of the first-neighbor pruning.
(b,c) The size $S$ and $R$ as a function of $p_H$ for the second-neighbor hyperedge pruning with $p_N=0.6$.
The numerical results (symbols) for $N=10^5$ and the analytical predictions (lines) are shown together 
for various core thresholds $(k,n)$ and anchor density $\theta$.
All hypergraphs have Poisson degree and cardinality distributions given by Eqs.~(\ref{eq:Poisson}) and (\ref{eq:Shifted Poisson}) with $\langle q \rangle=7.6$ and $\langle m \rangle=3.8$.
}
\label{fig:fig4}
\end{figure}

\begin{figure}
\includegraphics[width=\linewidth]{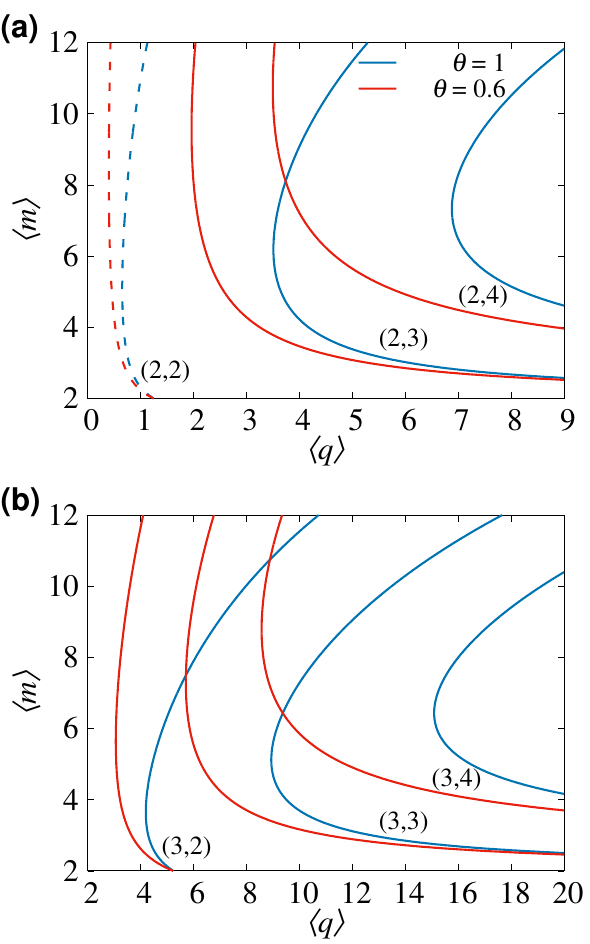}
\caption{
Phase diagram of second-neighbor hyperedge $(k,n)$-core percolation with anchor nodes on random Poisson hypergraphs in the $\langle q \rangle$-$\langle m \rangle$ plane. Here, $p_N=0.8$ and $p_H=1$.
The dashed lines and the solid lines denote continuous transitions and discontinuous transitions, respectively.
}
\label{fig:fig5}
\end{figure}

Figure~\ref{fig:fig4} compares the analytical predictions with numerical simulations for the second-neighbor pruning processes on random Poisson hypergraphs with $\langle q \rangle = 7.6$ and $\langle m \rangle = 3.8$. 
In Fig.~\ref{fig:fig4}(a), we show the fraction of hyperedges $R$ as a function of the node survival probability $p_N$ under the second-neighbor node pruning (with $p_H = 1$).
Comparison with the first-neighbor pruning is depicted by the dashed lines.
Since the fraction of nodes $S$ for the second-neighbor node pruning is identical to that of the first-neighbor pruning, we omit the comparison between the theory and numerical results. 
In addition, Figs.~\ref{fig:fig4}(b) and \ref{fig:fig4}(c) show the fractions of giant components $S$ and hyperedges $R$, respectively, as a function of the hyperedge survival probability $p_H$ under the second-neighbor hyperedge pruning (with fixed node damage $1-p_N = 0.4$). 
In all cases, the analytical solutions (lines) derived from the respective equations are in excellent agreement with the numerical results (symbols). Similar to the first-neighbor case, for $(k, n) = (2, 2)$, the transition shifts from continuous to discontinuous as $\theta$ increases, whereas for higher core thresholds such as $(2, 3)$ and $(3, 2)$, the transitions remain discontinuous for all $\theta$.

Figure~\ref{fig:fig5} shows the phase diagrams for the second-neighbor hyperedge pruning in the $\langle q \rangle$-$\langle m \rangle$ plane at $p_N = 0.8$ and $p_H = 1$. 
Here, we only present hyperedge pruning because the phase diagram of the second-neighbor node pruning is the same as that of the first-neighbor pruning.
In Fig.~\ref{fig:fig5}(a) with $k = 2$, the $(2, 2)$ phase boundary is composed of a continuous transition line (dashed) and a discontinuous transition line (solid). As the anchor probability $\theta$ increases from $0.6$ to $1.0$, the continuous transition region shrinks, and the transition boundaries shift toward higher average degree $\langle q \rangle$ and cardinality $\langle m \rangle$. For higher cardinality thresholds $n =3, 4$, the boundaries are purely discontinuous across all $\theta$. 
In Fig.~\ref{fig:fig5}(b) for $k = 3$, all transition lines for $n =2, 3, 4$ are only discontinuous regardless of $\theta$. 
These phase diagrams show that under the second-neighbor hyperedge pruning, increasing either the anchor density $\theta$ or the connectivity requirements $(k, n)$ consistently delays the percolation threshold, requiring a denser hypergraph for the giant $(k, n)$-core to appear.

\section{Conclusions}

In this work, we have introduced a $(k,n)$-core percolation model with anchor nodes on hypergraphs
in which the nodes of a hyperedge play different functional roles: the failure of an anchor node of a hyperedge dismantles the entire hyperedge, while the failure of the other nodes simply reduces the hyperedge size.  In our theoretical framework, each node of a hyperedge is one of its anchor nodes with probability $\theta$.
We have developed a comprehensive theoretical framework for $(k,n)$-core percolation models obeying  three distinct pruning processes: first-neighbor pruning and second-neighbor pruning acting either on nodes or on hyperedges. This formulation provides a unified framework in which the effect of  functional heterogeneity of the nodes is studied across different  ranges over which connectivity is ensured. The analytical predictions obtained from the corresponding self-consistency equations are in excellent agreement with numerical simulations on random hypergraphs.

Our results on the first-neighbor $(k,n)$-core percolation problem show that the presence of anchor nodes can qualitatively alter the robustness of higher-order networks. Specifically, by varying the value of $\theta$ our model can interpolate  between the factor graph $(k,n)$-core 
and the hypergraph $(k,n)$-core, which are recovered for $\theta=0$ and $\theta=1$, respectively \cite{lee2023k,bianconi2024nature}. 
Our results demonstrate that increasing the anchor density   increases  the connectivity required to sustain an extensive $(k,n)$-core and enlarges the non-percolating region of the phase diagram. More importantly, anchor nodes can change the nature of the percolation transition itself. For the $(2,2)$-core, increasing $\theta$ drives the system from a continuous transition to a discontinuous one, with the two regimes separated by a tricritical point. By contrast, for stronger core constraints, such as $(2,3)$ or $k>2$, the emergence of the core is generically discontinuous. Hence, functional dependence, captured by anchor nodes,  does not merely shift the percolation threshold but can transform a progressive degradation of the higher-order structure into an abrupt, catastrophic collapse.

The comparison between first- and second-neighbor pruning processes further shows that robustness depends not only on the structural organization of the hypergraph but also on the range of interactions that guarantees connectivity. In particular, the second-neighbor formulations distinguish situations in which only nodes or only hyperedges are recursively pruned, thereby capturing forms of extended-range connectivity that cannot be described by standard first-neighbor core percolation. These results indicate that different mechanisms ensuring connectivity beyond first neighbors can lead to extended notions of hypergraph robustness and different critical behavior.

The framework introduced here opens several directions for further research. A natural next step is to move beyond independently assigned anchor nodes and consider heterogeneous anchor probabilities depending on node degree, hyperedge cardinality, or other structural and functional properties. Correlations between anchor roles within and across hyperedges could be particularly relevant in real systems, where essential components are unlikely to be distributed independently. Another important direction is the investigation of targeted damage, in which anchor and non-anchor nodes have different failure probabilities, allowing one to quantify the vulnerability of higher-order systems to attacks on functionally critical components. More broadly, applying the theory to empirical hypergraphs \cite{mancastroppa2023hyper,mancastroppa2024structural} would make it possible to assess how the interplay between structural organization and heterogeneous functional roles shapes robustness in biological, technological, and social systems.

\begin{acknowledgments}
This research was supported in part by the National Research Foundation of Korea (NRF) grant funded by the Korea government (MSIT) (No. RS-2025-25433094), and by Global - Learning \& Academic research institution for Master’s · PhD students, and Postdocs (LAMP) Program of the National Research Foundation of Korea (NRF) grant funded by the Ministry of Education (No. RS-2024-00445180).
\end{acknowledgments}

\bibliography{percolation_bibliography}

\end{document}